\documentclass{svjour3}                     % onecolumn (standard format)
\smartqed  % flush right qed marks, e.g. at end of proof
\usepackage{graphicx}
 \usepackage{mathptmx}      % use Times fonts if available on your TeX system
\usepackage{color}
\definecolor{rblu}{rgb}{0.008,0.345,0.637}
\definecolor{darkgreen}{cmyk}{1,0,1,0.5}
\definecolor{darkblue}{cmyk}{1,1,0,0.35}
\definecolor{darkred}{rgb}{0.75,0.1,0.1}
\definecolor{darkorg}{rgb}{0.9,0.64,0.3}
\definecolor{YellowOrange}{rgb}{1,0.75,0.2}
\definecolor{gray}{rgb}{0.75, 0.75, 0.9}
\definecolor{orange}{rgb}{1.00, 0.64, 0.00}
\definecolor{dgrn}{cmyk}{1,0,1,0.5}
\definecolor{brown}{rgb}{0.55, 0.05, 0.05}
\begin{document}

\title{Exploring universality in nuclear clusters with Halo EFT
\footnote{Presented at the 21st European Conference on Few-Body Problems 
in Physics, Salamanca, Spain, 30 August - 3 September 2010}
\thanks{Work supported by the Dutch Stichting voor Fundamenteel
Onderzoek der Materie under programme 104.}
}


\author{R. Higa %\and
%        Second Author %etc.
}


\institute{R. Higa \at
              Kernfysisch Versneller Instituut, Rijksuniversiteit Groningen\\
              Zernikelaan 25, 9747 AA Groningen, The Netherlands\\
              \email{R.Higa@rug.nl}           %  \\
}

\date{Received: date / Accepted: date}

\maketitle

\begin{abstract}
I present results and highlight aspects of halo EFT to 
loosely bound systems composed of nucleons and alpha particles, with 
emphasis on Coulomb interactions. 
\end{abstract}

\section{Introduction}
\label{intro}
The physics of exotic nuclei still keeps motivating several research 
initiatives worldwide. 
Dedicated ongoing and future experiments promise to deliver more intense 
beams of rare isotopes along with new sophisticated detection techniques, 
paving the way to explore the limits of existence of several unknown 
nuclear systems and their unusual properties. 

Halo nuclei and nucleon-alpha clusters are particular examples of exotic 
nuclei and constitute the focus of this talk. They are normally characterized 
by a large structure relative to the typical size of each of its components, 
nucleons and/or stable nuclei. The large-distance physics of those systems 
is a response to the shallowness of their separation binding energies, 
$B_{lo}\sim 0.1$ MeV. To them is associated a low-momentum scale 
$M_{lo}\approx\sqrt{2\mu B_{lo}}$ 
which contrasts with a high-momentum $M_{hi}$ set by the 
energy required to excite a core, usually of the order of a few MeV. 
This separation of scales matches quite well with the ideas of 
effective field theories (EFTs), where the 
ratio of scales sets an expansion parameter that provides systematic 
and model-independent predictions, as well as more rigorous control 
over theoretical uncertainties. 

Halo/cluster EFT has been developed to account for certain 
aspects of loosely bound nuclear clusters, namely, low-energy resonances and 
Coulomb interactions. In the following I explain how these features are 
handled, with $\alpha\alpha$ and $p\alpha$ systems as examples. 

%%%%%%%%%%%%%%%%%%%%%%%%%%%%%%%%%%%%%%%%%%%%%%%%%%%%%%%%%%%%%%%%%%%%%%%%%%%%%%
\section{$\alpha\alpha$ and $p\alpha$ systems}
\label{sec:2}

The power counting for low-energy narrow resonances was developed in 
\cite{BHvK}. Unlike shallow bound states, a higher amount of fine-tuning 
is required to produce the expected energy dependence of the amplitude. 
That means a non-static two-body propagator or, equivalently, the sum 
of effective range contributions to all orders. As an example let us start 
with the $\alpha\alpha$ interaction, whose strong part is described by the 
following Lagrangian 
\begin{eqnarray}
{\cal L}&=&
\phi^{\dagger}\bigg[i\partial_0+\frac{\vec\nabla^2}{2m_{\alpha}}\bigg]\phi
-\,d^{\dagger}\bigg[i\partial_0+\frac{\vec\nabla^2}{4m_{\alpha}}-\Delta\bigg]d
%\nonumber\\&&
%
+g\,\big[d^{\dagger}\phi\phi+(\phi\phi)^{\dagger}d\big]
+\cdots\,.
\label{eq:LOLag}
\end{eqnarray}
We introduce an auxiliary (dimer) field $d$ with ``residual mass'' $\Delta$, 
carrying the quantum numbers of two alphas in $S$-wave and coupling with 
their fields via the coupling constant $g$. For a non-static $d$ propagator 
$\Delta$ has to scale as the kinetic energy, $\sim M_{lo}^2/4m_{\alpha}$, 
already two orders away from the natural scaling $\sim M_{hi}^2/4m_{\alpha}$.
The dots stand for higher order terms in a derivative expansion. 
Electromagnetic interactions are introduced via minimal substitution, with 
Coulomb forces being the dominant ones~\cite{HHvK}. The Sommerfeld parameter 
$\eta=Z_{\alpha}^2\alpha_{em}m_{\alpha}/2k=k_C/k$ sets the magnitude of 
Coulomb interactions, where $k_C$ is the inverse of the $\alpha\alpha $ Bohr 
radius. The fact that $k_C$ is numerically of $O(M_{hi})$ requires the sum 
of Coulomb photons to all orders. The technique to handle this problem was 
given by Kong and Ravndal, using established ideas of Coulomb Green's function 
and two-potential scattering~\cite{KR00}. The Coulomb-distorted strong 
amplitude acquires the form of a Coulomb-modified effective range expansion, 
with the shape parameter term treated as a perturbation~\cite{HHvK}. 

The $\alpha\alpha$ system is remarkable for having a scattering length of 
the order of 2000 fm, while the effective range and shape parameter obey 
natural dimensional analysis, $r_0\approx 1\mbox{ fm}\sim M_{hi}^{-1}$ and 
${\cal P}_0\approx 1.5\mbox{ fm}^3\sim M_{hi}^{-3}$. 
The $\alpha\alpha$ resonance energy of $E_R\simeq 92$ keV sets the 
low-momentum scale to $M_{lo}\sim\sqrt{m_{\alpha}E_R}\approx 20$ MeV, well 
below the high-momentum scale set by the pion mass or the excitation energy 
of the alpha particle, $M_{hi}\sim m_{\pi}\sim\sqrt{m_{\alpha}E_{\alpha}^{*}}
\approx 140$ MeV. With an expansion parameter around 1/7, the 
strong scattering length $a_{0,s}\sim m_{\alpha}g^2/\Delta\sim 
M_{hi}/M_{lo}^2$ can at most be of the order of few hundreds of MeV. 
We found in our study~\cite{HHvK} that most of the remaining scaling factor 
comes from a detailed cancellation between strong and electromagnetic 
interactions, an incredible amount of fine tuning. 
This is the outcome of an exponentially suppressed resonance width due to a 
large Coulomb barrier, entangled with the location of the resonance very 
close to the $\alpha\alpha$ threshold. 
This fine tuning was not expected and is not well understood, but leads to some 
interesting scenarios. First, an increase on the strong force by a few 
percent is enough to produce a bound ${}^{8}{\rm Be}$, which could have 
drastic astrophysical consequences. Second, the already large fine tuning in 
the strong parameters lead to an unitary amplitude at leading order (LO) when 
Coulomb is turned off. The ${}^{8}{\rm Be}$ would then be a bound 
state at zero energy, and the system with three $\alpha$'s would 
exhibit an exact Efimov spectrum~\cite{HHvK,BHrev}. Although Coulomb forces 
are highly non-perturbative, the fact that both ${}^{8}{\rm Be}$ and the 
Hoyle state (a ${}^{12}{\rm C}$ excited state $\sim 400$ keV above the 
$3\alpha$ threshold) remains very close to threshold supports such picture 
close to the unitary limit. The Hoyle state is essential to 
describe the correct abundance of ${}^{12}{\rm C}$ in the universe. Its 
existence is usually given as an example of large fine tuning in the 
parameters of the underlying theory~\cite{oberh}, but how that relates to the 
fine tuning in the $\alpha\alpha$ system is almost an unexplored subject. 
Our study provides a humble step to address this issue. 
\begin{table}
\caption{$\alpha\alpha$ effective range parameters.}
\label{tab:1}
\begin{tabular}{llll}
\hline\noalign{\smallskip}
& $a_0$ ($10^3$ fm) & $r_0$ (fm) & ${\cal P}_0$ (fm${}^{3}$) \\
\noalign{\smallskip}\hline\noalign{\smallskip}
LO & $-1.80$ & $1.083$ & --- \\
NLO & $-1.92\pm 0.09$ & $1.098\pm 0.005$ & $-1.46\pm 0.08$ \\
Rasche & $-1.65\pm 0.17$ & $1.084\pm 0.011$ & $-1.76\pm 0.22$ \\
\noalign{\smallskip}\hline
\end{tabular}
\end{table}

Despite this puzzling fine tuning, the associated power counting seems to 
be the correct one to describe the scattering data. At LO we have no free 
parameters, since we use the latest measurements of the resonance position 
and width~\cite{Wue92} as input. At NLO, the extra parameter is fitted to 
the $\alpha\alpha$ phase shifts. The large cancellation between strong and 
electromagnetic forces allows us to extract the effective range parameters 
with an accuracy better than previous determinations~\cite{rasche}. 

\begin{figure*}
  \includegraphics[width=0.42\textwidth]{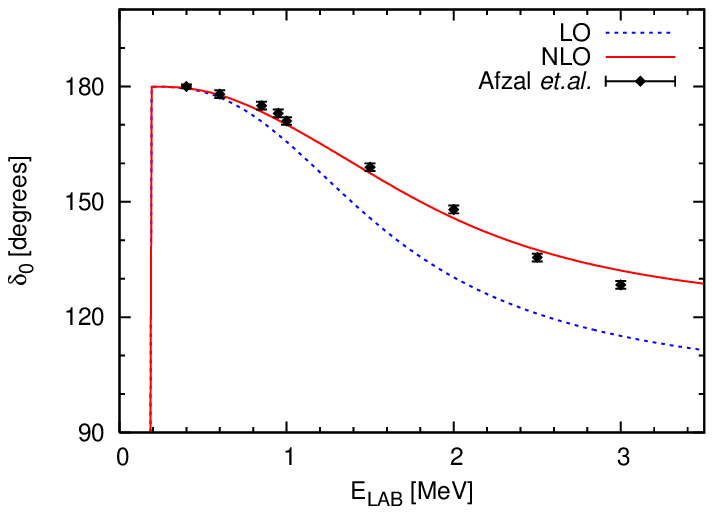}
  \includegraphics[width=0.42\textwidth]{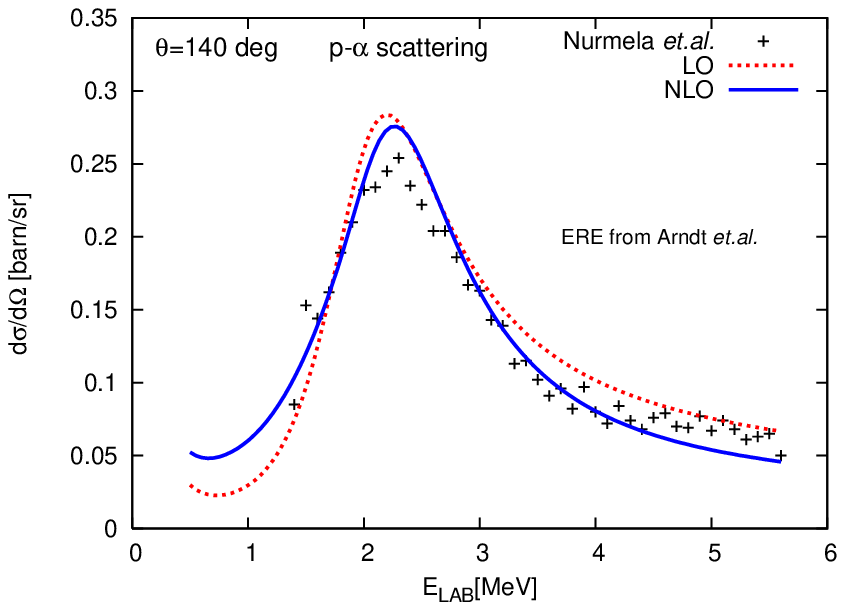}
\caption{$S$-wave $\alpha\alpha$ scattering phase shifts (left panel) and 
$p\alpha$ differential cross-section at $140^{\circ}$ CM angle (right panel).}
\label{fig:1}
\end{figure*}

The EFT approach to $p\alpha$ scattering follows steps similar to the 
$\alpha\alpha$ case~\cite{RBvK}. However, in this situation the envelope of 
the resonance is mainly given by the angular momentum barrier --- Coulomb 
forces provide just a correction to the strong part. At LO the amplitude 
receives contributions of both the $S_{1/2}$ and $P_{3/2}$ except around the 
$P_{3/2}$ resonance, which is enhanced. At NLO the $S_{1/2}$ effective range 
and $P_{3/2}$ shape parameter enter as perturbations. The $P_{1/2}$ partial 
wave contributes only at higher orders~\cite{BHvK,RBvK}. Preliminary result 
is shown in the right panel of Fig.~\ref{fig:1} using the effective range 
parameters from Arndt {\em et al.}~\cite{arndt73}, compared to the 
measurements performed by Nurmela {\em et al.}~\cite{nurmela}. 
The shape of the resonance is overall well reproduced in the cross-section. 
The small discrepancy at the resonance peak reflects the smaller values 
obtained by Ref.~\cite{nurmela,pusa} relative to previous 
measurements used in Ref.~\cite{arndt73}. 

\begin{acknowledgements}
I wish thank my collaborators and the organizers for this very interesting 
conference. 
\end{acknowledgements}



\end{document}